\documentclass{PoS}

\usepackage{amsmath}
\usepackage{amssymb}

\usepackage[version=3]{mhchem}	
\newcommand{\mbb}{m_{\beta \beta}} 
\newcommand{\ml}{m_\mathrm{lightest}} 

\newcommand{\el}{\text{e}}		\newcommand{\pr}{\text{p}}		\newcommand{\neu}{\text{n}}

\newcommand{\meV}{\text{meV}}				
\newcommand{\MeV}{\text{MeV}}	\newcommand{\GeV}{\text{GeV}}	

\newcommand{\SM}{SM}

\title{Testing creation of matter with neutrinoless double beta decay}

\ShortTitle{Testing creation of matter with neutrinoless double beta decay}

\author{{Stefano Dell'Oro}\\%
      Gran Sasso Science Institute, L'Aquila, Italy\\
      INFN - Laboratori Nazionali del Gran Sasso, Assergi (L'Aquila), Italy\\
      Center for Neutrino Physics, Virginia Polytechnic Institute and State University, 
			24061 Blacksburg, Virginia, USA\\
      E-mail: \email{stefano.delloro@gssi.infn.it}}

\author{{Simone Marcocci}\\%
      Gran Sasso Science Institute, L'Aquila, Italy\\
      Dipartimento di Fisica, Universit\`a degli Studi and INFN, Genua, Italy\\
     Fermi National Accelerator Laboratory, 60510 Batavia, Illinois, USA\\\
      E-mail: \email{simone.marcocci@gssi.infn.it}}

\author{\speaker{Francesco Vissani}\\%
      Gran Sasso Science Institute, L'Aquila, Italy\\
      INFN - Laboratori Nazionali del Gran Sasso, Assergi (L'Aquila), Italy\\
      E-mail: \email{francesco.vissani@lngs.infn.it}}

\abstract{In this brief review, the importance of the so called {\em neutrinoless double beta decay} transition in the search 
		for physics beyond the Standard Model is emphasized. 
		The expectations for the transition rate are examined in the assumption that ordinary neutrinos 
		have Majorana masses. 
		We stress the relevance of cosmological measurements and discuss the uncertainties implied by nuclear physics.
		This work is based on the review paper of Ref.~\cite{rev}.}

\FullConference{XVII International Workshop on Neutrino Telescopes\\
		 13-17 March 2017\\
		 Venezia, Italy}

\begin{document}

\section{Current understanding of what matter is made of}
\label{sec:mata}

	Let us begin by discussing what is matter or, more precisely, what are matter constituents. 
	A spontaneous answer -- {\em matter is made of atoms} -- is perfectly acceptable in normal conditions.
	However, at high energies or temperatures, new facts occur. Inside the Sun, matter is in the form 
	of plasma, i.\,e.\ electrons and atomic nuclei dissociate. Also, weak reactions such as 
	\begin{equation}
	\label{eq:pp}
		\pr + \pr \to \mbox{D} + \el^+ + \nu_\el
	\end{equation}
	produce, as net effects, 
	(1)~the disappearance of a proton accompanied by the appearance of a neutron inside the deuteron
	(2)~the disappearance of an electron from the medium, that annihilates with the newly formed positron, together with 
	the appearance of an electron neutrino.
	Therefore, physicists insist on considering neutrons and protons as the same form of heavy constituents of matter, 
	baryons, and electrons and neutrinos as the same form of light constituents of matter, leptons. 
	When neutrinos are included in the list, the net number of matter particles, either light or heavy, remains unchanged
	(antimatter particles count as $-1$).

	The features of the transformation in Eq.~\ref{eq:pp} are proper of all known reactions characterizing the 
	{\em Standard Model} (\SM)~of elementary particles and interactions, the best description that we have of matter and forces.
	Within this enormously successful theory, the number of leptons $L$ and the number of baryons 
	$B$ do not change in any perturbative reaction. The same conclusion applies to the baryon constituents, the 
	quarks, which carry $1/3$ of $B$ each. 
	Therefore, the \SM~definition of what is matter is simple: {\em matter is made of leptons and baryons (or quarks).}

	Now, let us examine a more subtle issue. 
	It is known -- to theorists, at least -- that the conservation of $B$ and $L$ is not exact. 
	At very high energies, e.\,g.\ in the early moments after the Big Bang, when the W$^-$ and Z$^0$ 
	bosons are in thermal equilibrium, $B$ and $L$ change due to non-perturbative transitions predicted by the \SM. 
	In fact, the baryon and lepton numbers are symmetries of the classical theory, but not of the full quantistic one.
	In these contexts, $B$ and $L$ are not conserved.
	
	The only exact symmetries of the \SM~are the combinations (differences)
	\begin{equation}
	\label{eq:fab4}
		B - L\ , \ L_\el - L_\mu\ , \ L_\mu - L_\tau\ , \ L_\tau - L_\el.
	\end{equation}
	Thus, we could have a transition where the energy converts into $3 \pr + \el + \mu + \tau$, 
	which may eventually turn out into 3 hydrogen atoms + 2 neutrino-antineutrino pairs. 
	This means that, in principle, the \SM~possesses the qualitative elements for the realization of the theoretical 
	program of Sakharov in order to explain the origin of the observed cosmic excess of atomic matter (the so called 
	Baryon Asymmetry of the Universe). 
	However, the program fails quantitatively: the \SM~produces an amount of baryons that is too small to be 
	relevant~\cite{misha}.

\section{Experimental tests of matter stability}
\label{sec:expa}

	There are various hints that the \SM~is not the ``final'' theory. 
	It is fair to stress that the strongest one is the observation of neutrino oscillations. 
	For example, the results of the T2K experiment~\cite{t2k} show that muon neutrinos become electron neutrinos during 
	their propagation, and therefore $L_\el - L_\tau$ and $ L_\mu-L_\tau$ are not exact symmetries of nature. 
	Together with the evidence of the tau neutrino appearance in a muon beam (result obtained by the OPERA 
	experiment~\cite{opera}), this proves that the only symmetry in Eq.~\ref{eq:fab4} that can be possibly respected
	is $B-L$.

	Once we realize that the \SM~is not infallible, we are led to search even more.
	It is of particular interest to test whether the conservation of the total number of baryons/leptons is verified 
	in every reaction or not. 
	The experimental investigations of such questions offer uniques chance of obtaining direct information on the 
	theory that extends the \SM, possibly leading to understand the origin of the matter.
	Some of these are considered particularly important and promising. 
	The first investigation is the search for the disintegration of nucleons into lighter particles. 
	Prototype processes of this kind are the decays
	\begin{align}
	\label{eq:zio1}
		\pr &\to \el^+ + \pi^0 \\ \neu &\to \el^- +  \mathrm{K}^+.\label{eq:zio15}
	\end{align}
	In particular, the former is actively searched.
	In general, there are many possible channels where energy and charge are respected.
	In the above examples, $B-L$ is conserved in the former process, but not in the latter.
	A transition where the (total and electron) lepton number is violated by two units is of the same class of importance.%
	\footnote{The assessment concerning  the ``importance'' takes into account, on the one hand, the relevance of the test 
		for the \SM\ and on the other one, the current chances of realizing a successful experiment.}
	In this case, a nucleus with mass number $A$ emits a pair of electrons, thus increasing its charge $Z$ by two units
	\begin{equation}
	\label{eq:zio2}
		(A,Z) \to (A,Z+2) + 2 \el^-.
	\end{equation}
	Another interesting test is the hypothetical oscillation
	of neutrons into antineutrons, which violates $B$ by two units. 
	
	Let us focus on the processes of Eqs.~\ref{eq:zio1} and \ref{eq:zio2}. 
	In the former, $B$ and $L$ decrease by one unit. 
	In the latter, $L$ increases by two units.
	By using the definitions introduced in the previous section, we can say that in Eq.~\ref{eq:zio1} matter is destroyed, 
	while in Eq.~\ref{eq:zio2} matter is created. 
	However, a deep property of relativistic quantum field theory, the ``crossing symmetry'',	inseparably links creation 
	and annihilation. Therefore, the difference between these processes is more superficial than it might appear.
	It is even more superficial in the just outlined context of the \SM, 
	where the violations of $L$ and $B$ have a very similar status and are tightly connected.

\paragraph{Issues with the traditional nomenclature}

	The above discussion shows the great importance of the processes of Eqs.~\ref{eq:zio1} and~\ref{eq:zio2} as tests 
	of the \SM.
	However, while the importance of the former is correctly reflected by the traditional denomination, namely
	{\em proton decay}, the traditional denomination of the latter, {\em neutrinoless double beta decay}, 
	hinders the recognition of this fact. 
	If we examine this second designation with a critical attitude, we note that:
	\begin{enumerate}
		\item the term `neutrinoless' sounds preposterous, since it proclaims that some particles (neutrinos) are 
			absent from the reaction.
			The transition is not classed by its own properties, but by something that it does {\em not} have.
		\item the definition of the transition is not self-contained, since it makes implicit reference to another transition, 
			i.\,e.\ the ordinary double beta decay (in which two neutrinos are emitted).
		\item the electron is indicated as `$\beta$-ray' following the terminology introduced by Rutherford and widely used 
			in nuclear physics, but much less in other fields of science, whereas the word `electron' 
			(known also to schoolchildren) would be equally good and appropriate.
	\end{enumerate}
	A more effective and precise description is anyway possible.
	The transition of Eq.~\ref{eq:zio2} could be simply called {\em creation of electrons}, specifying 
	the clause {\em`in a nuclear transformation'} if one wishes so.%
	\footnote{An alternative designation could include the suffix -genesis, from the Greek word for creation. 
		However, the term matter-genesis is too generic, the term electro-genesis is already used in physiology and biology 
		to describe the change of electric potential of a cell or the discharge of an electric eel, while the term 
		lepto-genesis is already used by theorists working in particle physics to denote a theory of the origin of the matter 
		(and also by copts in referring to a sacred book on the story of Mankind immediately after Adam and Eve).} 
	This expression is more precise and describes a positive quality of the reaction. 
	It stresses the relevance of the process in the context of the \SM~and not only. 
	When compared to `proton decay', we see that both expression concern 
	matter transformations, either destruction or creation.


\section{Majorana upgrade of the \SM}
\label{sec:maja}

	It is ironic that the most plausible extension of the \SM~suitable to explain why the differences of lepton numbers 
	are not respected (and potentially testable), was actually proposed already in 1937 by Ettore Majorana, 
	well before the \SM~was conceived.  
	Let us summarize the main features of this extension:
	\begin{itemize}
		\item in the \SM, neutrinos (antineutrinos) are strictly produced as left-handed (right-handed) particles. 
			Since neutrino masses are non-zero, this cannot be exactly true.
		\item the hypothesis of Majorana can be formulated as follows: in the rest frame, neutrinos and antineutrinos
			are just the same particle and are distinguished only by the spin.
		\item Majorana neutrinos are the only fermions to be matter and antimatter at the same time. 
			The difference between neutrinos and antineutrinos is not a Lorentz invariant concept, 
			and $L$ must be violated at the order $m_\nu/p_\nu$.
		\item since in usual conditions neutrinos are ultra-relativistic, so that $m_\nu \ll p_\nu$,  
			the deviation from this limit is not observable in most cases.
	\end{itemize}
	When we accept the attitude that the \SM~should be extended, we can recognize further specific merits of the 
	Majorana's hypothesis.
	By examining the motivations within the modern context of discussion, we could follow this way of thinking:
	\begin{enumerate}
		\item at ultra-high energy scales, e.\,g., at the $M_{\mbox{\tiny GUT}}$ scale where the gauge couplings unify,
			new higher dimensional operators, invariant under the \SM~symmetry, emerge from new physics.
			These operators are usually excluded from the \SM~by requiring the theory to be renormalizable.%
			\footnote{Notice that also Fermi interactions are non-renormalizable and are added to the quantum 
				electrodynamics in order to describe the existence of weak interactions.}
		\item the hypothesis of a high mass scale well agrees with experimental findings:
		\begin{equation}
			m_\nu \sim \frac{m_W^2}{M_{\mbox{\tiny GUT}}} = 65\,\meV  \times \frac{10^{14}\,\GeV}{M_{\mbox{\tiny GUT}}}
		\end{equation}
		\item the least suppressed is the (dimension-5) operator that describes the Majorana neutrino masses.
	\end{enumerate}
	
	An important consequence is that Majorana neutrinos violate the total lepton number. 
	Therefore they induce a contribution to the neutrinoless double beta decay (double electron creation) process. 
	This can be immediately visualized by Feynman diagrams, considering the core process 
	\begin{equation}
	\label{eq:WWee}
		W^- + W^- \to \el^- + \el^-
	\end{equation}
	where a virtual Majorana neutrino is exchanged.
	In ordinary condition it is more fair to talk in terms of nucleons. One might thus prefer to consider the transformation
	\begin{equation}
		2\neu \to 2\pr + 2\el^-
	\end{equation}
	induced by the process in Eq.~\ref{eq:WWee}, where virtual $W^-$s are radiated twice ($\neu \to \pr + W^*$),
	but nucleons are tightly bound in nuclei;
	an even more fair description needs to take this into account. 
	This is why it is even better to consider the form in Eq.~\ref{eq:zio2}, referring to transitions 
	such as \ce{^{76}Ge \to ^{76}Se + 2e^-}. 
	The latter description also suggests that, in order to get a reliable prediction, 
	we need an adequate understanding of the structure of the nuclei.

\section{Expectation on Majorana masses}
\label{sec:pi1}

	The Lagrangian density that describes Majorana masses is
	\begin{equation}
		\mathcal{L}_\mathrm{Majorana}=\sum_{\ell,\,\ell'\,=\,\el,\,\mu,\,\tau}
		\frac{m_{\nu_\ell \nu_{\ell'}}}{2} \, \nu_\ell^t(x) \, C^{-1} \, \nu_{\ell'}(x)~+~h.\,c.
	\end{equation}
	where $\nu_\ell$ is the quantized neutrino fields, while $C$ is the charge conjugation matrix. 
	The relevant parameter for the transition which is of interest for us is the ee-element of the $3\times 3$ mass matrix, 
	$m_{\rm \nu_\el\nu_\el}$. This appears evident from the fact that the electronic lepton number is violated twice. 
	The absolute value of this element is often named $\mbb$.
	
	It is possible to express the mass matrix in terms of the oscillation mixing matrix, 
	the (non-negative) eigenvalues $m_{\nu_j}$ (i.\,e.\ the masses) and a set of phases $\xi_j$ 
	called the Majorana phases:
	\begin{equation}
		m_{\nu_\ell \nu_{\ell'}} = U_{\ell j}^* \  U_{{\ell'} j}^* \   \el^{-i\,\xi_j}\   m_{\nu_j}
	\end{equation} 
	where the sum is over $j=1,2,3$.
	Unfortunately, the absolute mass and the Majorana phases are {\em not} probed by oscillations, the only evidence we 
	have to date of massive neutrinos.
	Therefore, the parameter $|m_{\rm \nu_\el\nu_\el}|$ 
	($\equiv \mbb$) is not known precisely. 

	It is customary to plot $|m_{\rm \nu_\el\nu_\el}|$ as a function of the lightest neutrino mass, $\ml$, 
	varying freely the Majorana phases $\xi_j$ and within the errors the other parameters~\cite{vis,Dell'Oro:2014yca}.
	Alternatively, the sum of the masses $\Sigma \equiv m_1+m_2+m_3$ can be used instead of $\ml$.
	This is advantageous, since $\Sigma$  	is probed by analyses of cosmological observations and  	
	since several works from cosmology, that appeared in the last couple of years~\cite{bar,mat},
	have stringently constrained $\Sigma$. The tight upper limit on $\Sigma$ implies 
	very small values of $|m_{\rm \nu_e\nu_e}|$, at most of the order of the value  
	$\sqrt{\Delta m^2_\mathrm{atm}} \simeq 50\,\meV$ pointed out the atmospheric neutrino oscillations, 
	see e.\,g.\ Ref.~\cite{rev} as a review.

	Regarding possible indications on the neutrino mass coming from theory, caution is necessary.
	At present, it is not possible to state that we succeeded in predicting the various parameters of massive neutrinos 
	actually measured, or even that we understood the masses of the charged fermions. 
	As a purpose of illustration, we mention here a hypothetical scheme (not a full fledged model) for the neutrino masses. 
	This is based on the simple consideration that the very large atmospheric neutrino mixing suggests that 
	the `$\mu-\tau$' block of the matrix contains elements of similar size, while the other ones are 
	smaller~\cite{ber-ros96,vis98}.
	This suggests that the neutrino mass matrix is proportional to
	\begin{equation}
	\label{stobarbon}
			\left(
			\begin{array}{ccc}
				\varepsilon  & \varepsilon  & \varepsilon  \\
				\varepsilon  & 1 & 1 \\
				\varepsilon  & 1 & 1 \\
			\end{array}
			\right)
		\quad \mbox{or} \quad
			\left(
			\begin{array}{ccc}
				\varepsilon^2  & \varepsilon  & \varepsilon  \\
				\varepsilon  & 1 & 1 \\
				\varepsilon  & 1 & 1 \\
			\end{array}
			\right)
	\end{equation}
	where the order parameter $\varepsilon$ is $<1$, while the overall mass scale is not predicted.%
	\footnote{Only the ordering in powers of $\varepsilon$ is indicated, so that there are also $\mathcal{O}(1)$ coefficients.}
	The former case is called ``extended dominant block'', while the latter ``electronic selection rule''.

	The expectations corresponding to these sets of mass matrices can be explored with random number generators for the  
	$\mathcal{O}(1)$ coefficients. 
	It was shown in Ref.~\cite{vis01} that the value $\varepsilon = \theta_\mathrm{C} \simeq 13^\circ$ 
	or $=\sqrt{m_\mu/m_\tau} \simeq 14^\circ$ 
	well agrees with the data.
	It prefers the solution of ``large mixing angle'' now confirmed, giving a large value of $\theta_{13}$ and 
	choosing the normal mass hierarchy case.
	In this case, we get the indication
	\begin{equation}
		|m_{\nu_\el\nu_\el}| = \mathcal{O}(1)\times \sqrt{\Delta m^2_\mathrm{atm}}\times \theta_\mathrm{C}^n
	\end{equation}
	where $n=1$ or $2$ in the cases of Eq.~\ref{stobarbon}.
	In particular, the latter case has smaller $|m_{\rm \nu_\el\nu_\el}|$
	but it is consistent with a U(1) selection rule.
	
	It is also possible to random-generate the values of the Majorana phases~\cite{benato,etal1,etal2}. 
	In this case, one typically expects largish values of $|m_{\rm \nu_e\nu_e}|$. 
	The hypothesis that ensures this outcome is a specific choice of the parameters 
	extracted with ``flat priors'', i.\,e.\ an assumed preference for the Majorana phases. 
	From a theoretical point of view, it seems easier to motivate a preference for  
	$|m_{\nu_\el\nu_\el}|$, since this quantity is proportional to the one that enters the Lagrangian density.

\section{From Majorana mass to half-life: some considerations}
\label{sec:pi2}

	The experiments can probe the half-life time of the decay of Eq.~\ref{eq:zio2}.
	We can express this quantity as 
	\begin{equation}
	\label{eq:half-life}
		\left[t_{1/2} \right] ^{-1} = G \cdot \mathcal{M}_\mathrm{nucl}^2 \cdot 
		\left| \frac{m_{\nu_\el\nu_\el}}{m_\el} \right|^2 
	\end{equation}
	where $G$ is a phase space that also describes the physics of the emitted electrons, 
	$\mathcal{M}_\mathrm{nucl} $ is the Nuclear Matrix Element (NME), which describes the transformation 
	of the nucleons involved in the transition and $ m_{\rm \nu_e\nu_e}$ is the ee-element of the Majorana 
	neutrino mass matrix in the basis where the charged leptons are diagonal.
	The electron mass $m_\el$ is traditionally chosen for purposes of normalization.
	It is worth to notice that in the case of a background-free experiment, the sensitivity on the half-life 
	improves proportionally to the collected {\em exposure}, i.\,e.\ the data taking time $\times$ the detector mass.

	This transition takes place inside nuclei and the momentum of the virtual nucleon is large, 
	$Q \sim \mathcal O (100\,\MeV)$ (inverse of the nucleonic size).
	It is thus much larger than the neutrino mass.
	Also, the axial coupling of the nucleons, $g_A$, is of great importance, since the decay rate scales approximatively 
	as $g_A^4$.

	Eq.~\ref{eq:half-life} shows that theory plays a fundamental role in extracting the information on the 
	neutrino mass. Therefore, it is important to discuss the uncertainties of the involved quantities and, in particular,
	to try to assess those concerning the NMEs.
	About 15 years ago, the uncertainties on the NME were claimed to be large, a factor of a few~\cite{feruglio}.
	More recently, the situation has improved, with different models providing results that agree within some tens of percent.
	In particular, the calculations by `QRPA' and `IBM-2' stay within $\simeq 30\%$ for most of the nuclei of interest. 
	
	This fact leads sometimes to speculate that the actual uncertainties on the NMEs are of this order. 
	However, it has to be noticed that the discrepancies with other results obtained with different methods of calculations 
	(in particular those based on the `shell model') suggest a more conservative approach. 
	Moreover, the same IBM-2 and QRPA models overestimate the NMEs much more than $30\%$ when predicting the rates of 
	the weak processes we can probe, e.\,g.\ regular double beta decay (with emission of neutrinos) 
	and ordinary beta decay and electron capture~\cite{fas,barea}. 

	These simple considerations hint that maybe there is a systematic overestimation of the calculated NME, 
	and maybe there is also a common cause. 
	One attempt in this direction invokes the idea that the nuclear medium plays a role. 
	A well-known proposal of this type is the one discussed in Refs.~\cite{barea,Barea:2015kwa}, 
	within the frame of the IBM-2 model,
	where it is assumed that the couplings do not have the same value in free space and in the nuclei. 
	In the latter case, these are actually \emph{quenched}, indeed due to the presence of the nuclear medium 
	(hence the need of a ``renormalization'').
	By using the IBM-2 calculations for the regular double beta decay, $g_A$ has to be be empirically adjusted with 
	a trend that decreases with the atomic number $A$ approximatively as $g_A=1.269 \cdot A^{-0.18}$
	in order to match the experimental findings.
	
	On the other side, the quenching of the axial coupling constant could turn out to be a cursory way to account for the 
	incompleteness of the present calculations. 
	Anyway, we think that, to date, caution would recommend to vary $g_A$ in a rather wide range to remind 
	that the uncertainties on the NMEs are unlikely to be small.

	The issue of the quenching/renormalization of $g_A$ should not be considered as a theory. 
	However, if there is a 
	physical cause for this effect, this is likely to depend upon the momentum $Q$, 
	since at very high $Q$ nucleons can be treated as free particles and 
	free nucleons do not suffer any quenching.
	And maybe there is only a loose connection between the two processes of double electron emission 
	when two or no neutrinos are emitted: In fact, the transferred momentum is quite different.
	When $Q$ is larger, as in the case of no neutrino emission in which we are interested, 
	$g_A$ could be closer to the free nucleon value ($\simeq 1.269$), 
	or to that of quark matter ($=1$)~\cite{mgs,em}.

\color{black}
\section{Summary and discussion}

	The occurrence of neutrino oscillations requires an upgrade of the \SM. 
	The process of creation of electrons in a nuclear transition (alias neutrinoless double beta decay) 
	tests matter stability and it is important as proton decay in this context
	(Secs.~\ref{sec:mata} and~\ref{sec:expa}).

	Majorana neutrinos are matter and antimatter at the same time (Sec.~\ref{sec:maja}). 
	They can induce the creation of electrons in a nuclear transition, with a rate linked to oscillations and to 
	cosmological measurements (Sec.~\ref{sec:pi1}).

	Uncertainties on this process rate are large. This is mostly due to particle physics, but 
	a large contribution also comes from nuclear physics (Sec.~\ref{sec:pi2}).

\medskip

	The present scenario in this field is dynamic and in continuous evolution, with very large experimental efforts.
	We are beginning to probe a very interesting region of the neutrino mass spectrum, 
	but, even for ``background-free'' conditions, the detector mass can represent an issue.
	Compare these considerations with those from the interventions of 
	G\'omez-Cadenas, Pozzi, Shirai, Sch\"onert and Wilkerson at this conference and 
	see again Ref.~\cite{rev} and in particular Table~X there, where the possible need of multi-ton experiments is pointed out.

\medskip
	Finally, let us recall that Majorana neutrino masses are a very reasonable source of $L$-violation, but we want to 
	measure the half-life time of a process when no neutrino is emitted in the final state. 
	The connection with neutrino masses is plausible but not proved. 
	If new sources of $L$-violation at low energy (TeV?) exist, surprises are possible. 


\subsubsection*{Acknowledgments}

	F.V.\ thanks M.~Mezzetto for the invitation at {\em Neutrino Telescopes 2017} and A.~Gallo Rosso for precious discussions.



	
\end{document}